\documentclass[10pt,twocolumn,letterpaper]{article}

\usepackage[pagenumbers]{wacv} % To force page numbers, e.g. for an arXiv version

\definecolor{wacvblue}{rgb}{0.21,0.49,0.74}
\usepackage[pagebackref,breaklinks,colorlinks,allcolors=wacvblue]{hyperref}
\usepackage{bm}
\usepackage{amsfonts}
\usepackage{amsmath}
\usepackage{float}
\usepackage{caption}

\def\wacvPaperID{761} % *** Enter the WACV Paper ID here
\def\confName{WACV}
\def\confYear{2026}

\title{ARGS: Advanced Regularization on Aligning Gaussians over the Surface}

\author{Jeong Uk Lee\\
KAIST\\
{\tt\small jeong19@kaist.ac.kr}
\and
Sung Hee Choi\\
KAIST\\
{\tt\small sunghee@kaist.edu}
}

\begin{document}
\maketitle
\begin{abstract}
Reconstructing high-quality 3D meshes and visuals from 3D Gaussian Splatting(3DGS) still remains a central challenge in computer graphics. Although existing models such as SuGaR offer effective solutions for rendering, there is is still room to improve improve both visual fidelity and scene consistency. This work builds upon SuGaR by introducing two complementary regularization strategies that address common limitations in both the shape of individual Gaussians and the coherence of the overall surface. The first strategy introduces an effective rank regularization, motivated by recent studies on Gaussian primitive structures. This regularization discourages extreme anisotropy—specifically, "needle-like" shapes—by favoring more balanced, "disk-like" forms that are better suited for stable surface reconstruction. The second strategy integrates a neural Signed Distance Function (SDF) into the optimization process. The SDF is regularized with an Eikonal loss to maintain proper distance properties and provides a continuous global surface prior, guiding Gaussians toward better alignment with the underlying geometry. These two regularizations aim to improve both the fidelity of individual Gaussian primitives and their collective surface behavior. The final model can make more accurate and coherent visuals from 3DGS data.
\end{abstract}
    
\section{Introduction}
\label{sec:intro}

The ability to reconstruct detailed 3D models from collections of 2D images has long been a central pursuit in computer vision and graphics. Recently, 3DGS~\cite{author1} has emerged as a groundbreaking technique, achieving real-time rendering speeds and state-of-the-art quality for novel view synthesis. By representing scenes with millions of explicit 3D Gaussian primitives and employing a differentiable tile-based rasterizer, 3DGS bypasses the costly neural network queries inherent in methods like Neural Radiance Fields(NeRFs)~\cite{author2}, thus offering significant advantages in training and rendering efficiency. This advancement has accelerated considerable interest in various applications, from virtual and augmented reality to interactive content creation.

Despite its success in rendering, converting the optimized, and often unstructured, set of 3D Gaussians into high-quality, explicit surface meshes—a preferred representation for many downstream graphics applications like editing, simulation, and animation—remains a significant challenge. The optimization process in 3DGS can result in Gaussians becoming unorganized or adopting highly anisotropic shapes~\cite{author4} where one variance dominates the others. Such configurations can lead to visual artifacts and suboptimal geometric details in novel views or areas with sparse input data.

To address these issues, SuGaR(Surface-Aligned Gaussian Splatting)~\cite{author3} was introduced as a method for efficient 3D mesh reconstruction and high-quality rendering from 3DGS representations. SuGaR employs a regularization term to encourage Gaussians to align with scene surfaces, extracts a mesh using Poisson reconstruction, and then binds new Gaussians to this mesh for refined rendering and editability. While SuGaR significantly improves mesh extraction from 3DGS, the geometric conditioning of individual Gaussians and the global coherence of the represented surface offer avenues for further enhancement.

This paper proposes two synergistic enhancements to the SuGaR framework, aiming to improve both the fidelity of individual Gaussian primitives and the overall consistency of the reconstructed surface. Building on the effective rank regularization introduced by \cite{author4}, our method is able to represent local patches well with respect to the object surfaces. This term directly analyzes the covariance structure of each Gaussian, penalizing the formation of overly elongated, needle-like shapes and encouraging more voluminous, disk-like primitives. Such well-conditioned Gaussians are more robust for representing local surface and can reduce geometric artifacts.

Our second regularization incorporates a co-optimized neural Signed Distance Function(SDF) to serve as a global surface prior. Inspired by advancements in neural implicit surface learning, such as NeuS~\cite{author5} or VolSDF~\cite{author10}, we jointly train a small multi-layer perceptron (MLP) to represent an SDF of the scene alongside the Gaussian optimization. This neural SDF is regularized by an Eikonal term to ensure it maintains valid distance field properties. A novel loss component then encourages the 3D Gaussians to concentrate near the zero-level set of this dynamically learned SDF, promoting a more globally coherent and continuous surface representation across the entire scene.

By combining these principled regularization strategies, our enhanced SuGaR methodology aims to achieve significantly improved geometric accuracy, reduce surface noise and artifacts, and produce more detailed and consistent mesh rendering from 3D Gaussian Splatting. This work seeks to further bridge the gap between the efficient rendering capabilities of 3DGS and the demand for high-quality, structured 3D assets in modern graphics pipelines.

\section{Related Work}
\label{sec:formatting}

\begin{figure*}
  \centering
  \includegraphics[width=0.8\linewidth]{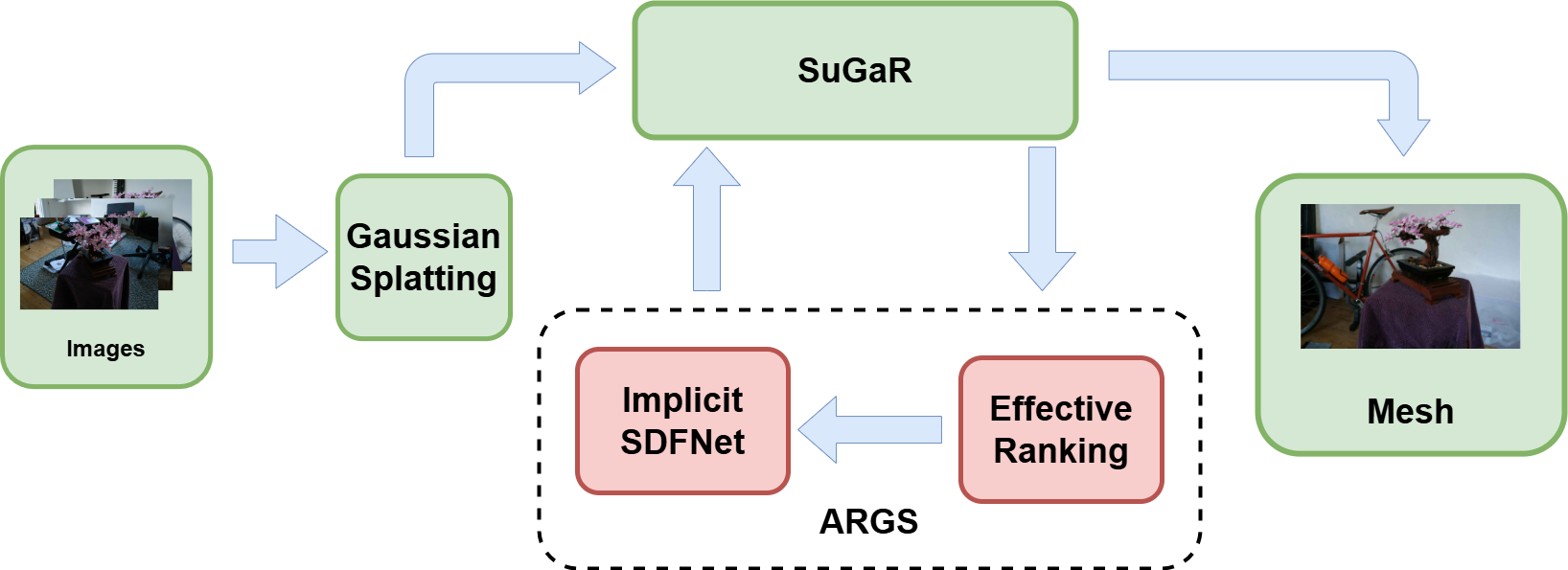}

   \caption{The general pipeline of our model. The images are processed in 3DGS and SuGaR to create mesh. During the process, we incorporate additional regularization to better match the gaussians to the surface.}
   \label{fig:onecol}
\end{figure*}

Our work is built upon recent advancements in 3DGS and methods for mesh conversion and geometric regularization, as well as techniques for neural implicit surface representation. We situate our contributions within these areas.

%-------------------------------------------------------------------------
\subsection{Mesh Conversion from Neural Volumetric Representations}

Obtaining high-quality, explicit surface meshes from neural volumetric representations like 3D Gaussian Splatting or Neural Radiance Fields is crucial for many downstream applications but presents distinct challenges.

For 3DGS, which achieves real-time rendering with explicit primitives, the primary challenge lies in interpreting a surface from a collection of potentially unorganized Gaussians~\cite{author1}. SuGaR, the foundation of our work, addresses this by regularizing Gaussians for surface alignment and then extracting a mesh via Poisson reconstruction from a derived density field~\cite{author3}. Other 3DGS-focused methods for geometry improvement or meshing include 2DGS~\cite{author6}, which uses 2D Gaussian disks for view-consistent geometry, GOF (Gaussian Opacity Fields)~\cite{author7}, which employs ray-Gaussian intersection for density estimation and regularization, and NeuSG~\cite{author8}, which combines 3DGS with neural implicit models for surface refinement. GS2Mesh~\cite{author13} uses stereo image pairs to estimate depth maps and get Truncated Signed Distance Function (TSDF) volume. Then it uses Marching Cube algorithm to extract mesh.

For NeRFs, which learn a continuous volumetric density and radiance, extracting accurate geometry is difficult because NeRFs are optimized for novel view synthesis, often resulting in fuzzy or noisy geometry when a simple density threshold is used. To overcome this, NeRFMeshing~\cite{author12} proposes distilling a pre-trained NeRF into a Truncated Signed Distance Field represented by a Signed Surface Approximation Network(SSAN). The SSAN is supervised using depth percentile estimates rendered from the NeRF to approximate points inside, on, and outside the surface, respectively. A mesh is then extracted from this TSDF using Marching Cubes. NeRF2Mesh~\cite{author14} distill the geometry encoded in a trained NeRF into a SDF. Eikonal Loss is applied to the SDF to enforce it to behave like a true distance function. Then Marching cube is used for mesh extraction.

%-------------------------------------------------------------------------
\subsection{Regularization of 3D Gaussian Primitives}

The rendering quality of meshes derived from 3DGS heavily relies on the shape and disposition of the Gaussian primitives. Unconstrained, these primitives can adopt undesirable anisotropic forms, such as "needle-like" shapes, leading to artifacts~\cite{author4}.

SuGaR's original regularization encourages surface alignment by matching distance values derived from Gaussians to depth-based cues. This is an effective approach but may not fully prevent all degeneracies in individual Gaussian shapes. The Effective Rank Regularization, which forms one of our key enhancements, directly addresses this by analyzing the covariance of each Gaussian. Along with Effective Rank regularization, FreGS~\cite{author15} uses Total Variation(TV) regularization which encourages smoothness in Gaussian positions and features to promote spatial coherence.

%-------------------------------------------------------------------------
\subsection{Neural Implicit Surfaces and SDF-based Reconstruction}

Neural implicit representations, especially SDF, have proven powerful for high-fidelity 3D surface reconstruction~\cite{author5, author10}. An SDF defines a surface as its zero-level set, $f(x)=0$, and ideally satisfies the Eikonal constraint, $\left\| \bm{\nabla f(x)} \right\|_2 = 1$, for true distance properties.

Several methods learn neural SDFs from multi-view images. IDR (Multiview Neural Surface Reconstruction)~\cite{author9} uses differentiable surface rendering but often requires masks and can face optimization challenges with complex scenes.

To improve robustness, recent works combine SDF with volume rendering principles. ~\cite{author5} introduces a volume rendering scheme based on an SDF-derived density, enabling mask-free training and better handling of complex geometry. ~\cite{author10} proposes modeling the volume density directly as a function of the SDF, specifically using Laplace's Cumulative Distribution Function(CDF) applied to the SDF values. This formulation is argued to provide a beneficial inductive bias, allow for accurate ray sampling via opacity error bounds, and facilitate shape/appearance disentanglement. Both NeuS and VolSDF emphasize the Eikonal loss for regularizing the learned SDF. Other related works include UNISURF ~\cite{author11}, which learns an occupancy field via volume rendering.

As mentioned, ~\cite{author12} also leverages a learned implicit representation by distilling it from a pre-trained NeRF using depth cues derived from the NeRF's volume rendering. This implicit representation then allows for standard mesh extraction.

Our work is inspired by the success of these methods in achieving high-quality surfaces rendering via learned SDFs. We propose to integrate a co-optimized neural SDF, similar in spirit to those in NeuS or VolSDF but learned in conjunction with explicit Gaussian primitives, to serve as a global surface prior within the SuGaR framework.

%-------------------------------------------------------------------------
\subsection{Positioning the Proposed Work}

Our method enhances the SuGaR framework through two primary, synergistic regularization strategies. The effective rank regularization targets the local geometric quality of individual Gaussians, promoting well-conditioned, disk-like shapes. Simultaneously, the co-optimized neural SDF, inspired by principles from implicit surface learning methods such as NeuS and VolSDF. This dual sided learning can be found in many other works. For example, methods like GSurf ~\cite{author16} learns a SDF that is directly supervised by the 3D Gaussian primitives. This approach aims to tightly couple the two representations by training them concurrently. GSDF ~\cite{author17} employs a novel dual-branch structure where 3DGS branch offers efficient rendering, while the SDF branch manages a neural Signed Distance Field, focused on accurate surface reconstruction. This dual strategy aims to improve both the intrinsic quality of the Gaussian primitives and their collective adherence to a coherent surface.

%-------------------------------------------------------------------------

\section{Preliminary}

\begin{figure*}
  \centering
  \includegraphics[width=0.8\linewidth]{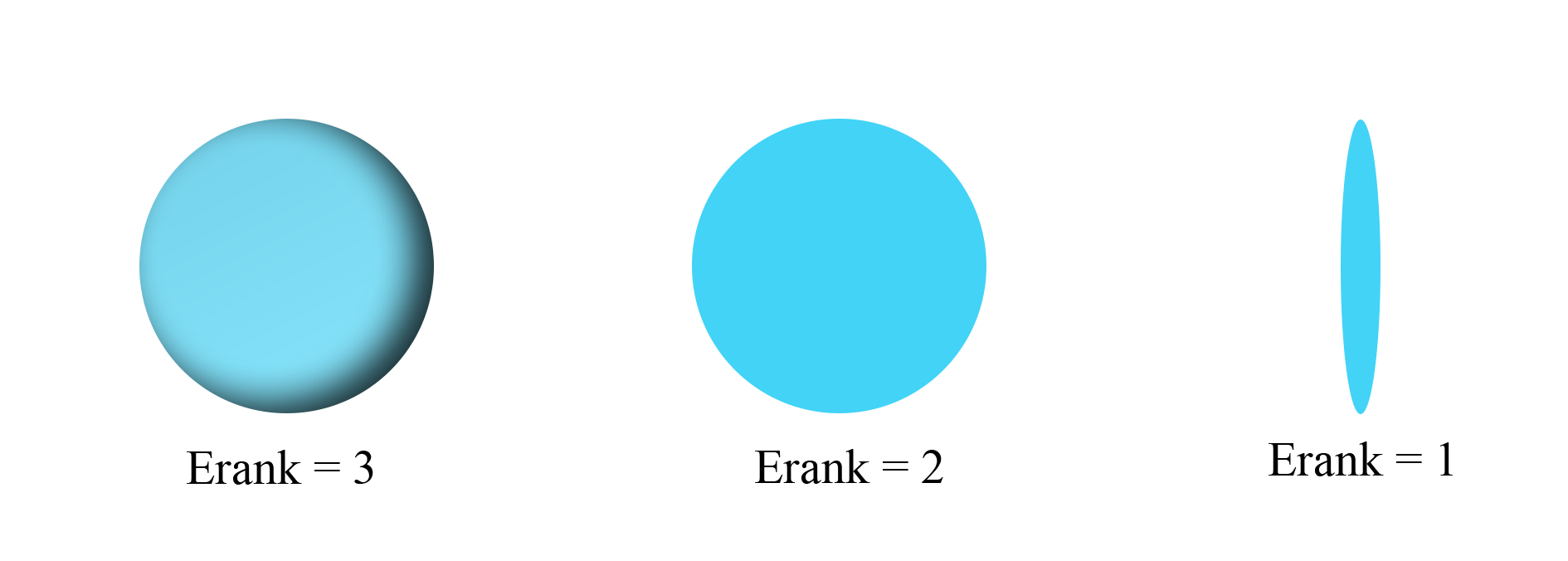}

   \caption{In the original Gaussian Splatting, many Gaussians degenerate into ”needle-like” shape. Also we prefer Gaussians to align to the surface. For these reasons, we optimize Gaussians to have effective rank of 2.}
   \label{fig:twocol}
\end{figure*}

\label{sec:prelim}
In this section, we introduce the concept of effective rank regularization from ~\cite{author4}, a key component for our method's ability to create well-conditioned Gaussians that represent local surface patches.
The effective rank $\operatorname{erank}(\mathcal{G}_K)$ is computed from its 3D scaling factors $\bm{s}_k = (s_{k1}, s_{k2}, s_{k3})$. We first need to obtain normalized squared scales.
\begin{equation}
\mathbf{q} = (q_1, q_2, q_3) = \left( \frac{s_1^2}{S}, \frac{s_2^2}{S}, \frac{s_3^2}{S} \right), \quad S = \sum_{i=1}^{3} s_i^2
\label{eq:two}
\end{equation}

\noindent
Then we need to take Shannon entropy of these terms.
\begin{equation}
\bm{H}(\mathcal{G}_k) = \bm{H}(q_1, q_2, q_3) = -\sum_{i=1}^{3} q_i \log q_i
\label{eq:three}
\end{equation}

\noindent
We can obtain the effective rank of Gaussians by then taking its exponent. 
\begin{equation}
\operatorname{erank}(\mathcal{G}_K) = \exp\{ H(\mathcal{G}_k) \}
\label{eq:four}
\end{equation}

A penalty term is applied, which sharply increases as $\operatorname{erank}(\mathcal{G}_K)$ approaches to 1. This term is formulated as $\max\left( -\log(\operatorname{erank}(\mathcal{G}_k) - 1 + \epsilon),\, 0 \right)$, where $\epsilon$ is a small constant for numerical stability.
The smallest scaling factor $\bm{s}_{k3}$ of the Gaussian is added to this penalty. The final per-Gaussian loss can be written as:

\begin{equation}
\mathcal{L}_{eff} = \sum_{k} \max\left( -\log(\operatorname{erank}(\mathcal{G}_k) - 1 + \epsilon),\, 0 \right) + s_{k3}
\label{eq:five}
\end{equation}

\section{Method}

\subsection{Overall Pipeline}

Our approach enhances the SuGaR framework\cite{author3} for 3D mesh conversion and rendering from 3DGS. We introduce two primary regularization strategies during SuGaR's coarse optimization phase: an effective rank regularization to refine the local geometry of individual Gaussian primitives and a co-optimized neural Signed Distance Function to establish a global surface prior. These components work synergistically to improve the overall look and coherence of the mesh.

The core of our method integrates these novel regularizations into the alignment stage of gaussians of SuGaR. The pipeline generally proceeds as follows:
\\
\begin{enumerate}
  \item \textbf{Optimization:} The process begins with a 3DGS model that has performed the initial optimization, typically from a sparse point cloud derived from Structure-from-Motion.
  \\
  \item \textbf{Coarse SuGaR Optimization with Novel Regularizers:} This stage is crucial for further optimization of the Gaussian primitives. Alongside SuGaR's original surface alignment objectives, our two key regularization components are applied. First the effective rank regularization is introduced to improve the shape of individual Gaussians. Second, a neural SDF network is jointly trained, and associated losses (Eikonal and SDF-Gaussian consistency) are applied to enforce a global, continuous surface structure. The loss function that is then applied can be written as:
  \begin{multline}
    \mathcal{L} = \mathcal{L}_{sugar} + \lambda_{SDF-Gauss} \mathcal{L}_{eff} \\
    + \lambda_{2} \mathcal{L}_{Eikonal} + \lambda_{3} \mathcal{L}_{SDF-Gauss}
    \label{eq:one}
    \end{multline}
  \item \textbf{Mesh Exteraction:} After the coarse optimization, a 3D mesh is extracted from the regularized Gaussian representation using SuGaR's established procedure, which involves sampling points on a density level set and applying Poisson reconstruction.
  \\
  \item \textbf{Further Refinement:} The extracted mesh can subsequently be used for SuGaR's refinement stage, where new Gaussians are bound to the mesh and further optimized. Our primary contributions focus on enhancing the coarse optimization to yield a superior mesh rendering quality.
  \\
\end{enumerate}
These regularizations are activated after an initial number of training iterations to allow the Gaussians to settle roughly in their place. The overall pipeline of the model is displayed in \cref{fig:onecol}.

\begin{figure*}
  \centering
  \includegraphics[width=0.8\linewidth]{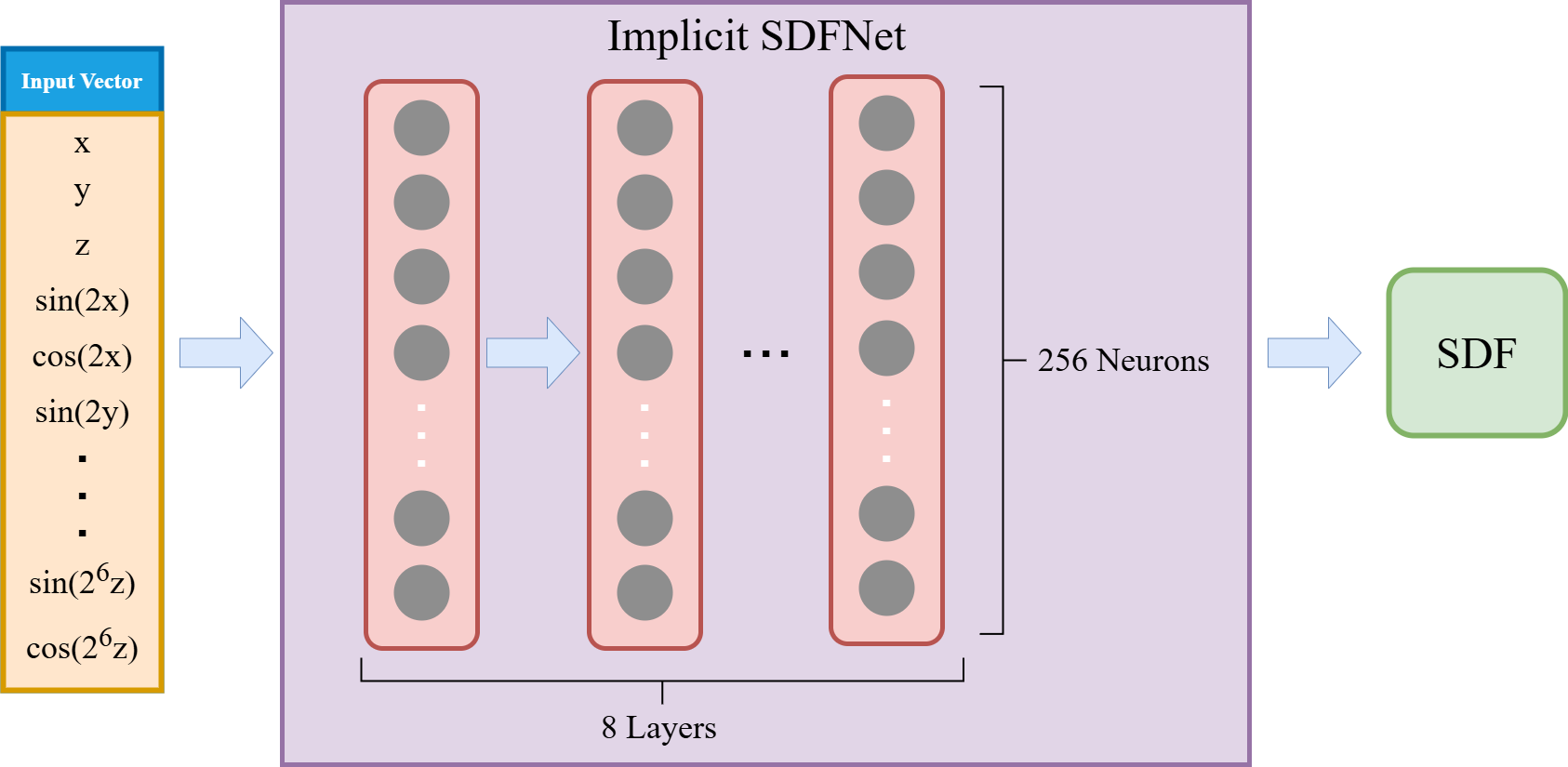}

   \caption{The Implicit SDFNet is comprised of 8 hidden layers with 256 neurons per each layer. Input vector with positional encoding in fed into the network to output a single SDF value.}
   \label{fig:threecol}
\end{figure*}

\subsection{Effective Rank Regularization for Gaussian Primitives}
Following the method proposed by Hyung et al.~\cite{author4}, we incorporate an effective rank regularization loss to prevent 3D Gaussians from degenerating into undesirable 'needle-like' shapes (effective rank $\approx 1$). This loss encourages each Gaussian primitive to adopt a more geometrically stable, 'disk-like' shape (effective rank $\approx 2$), which is better suited for representing surface patches. The visualization of effective rank is shown in \cref{fig:twocol}.
The effective rank is computed from the Gaussian's 3D scaling factors. The per-Gaussian loss, adapted from~\cite{author4}, is then formulated as \cref{eq:five}.
This loss is averaged over all Gaussians and added to the main training objective, weighted by a hyperparameter $\lambda_{erank}$.

\subsection{Implicit SDF Constraint for Global Surface Coherence}
To provide a strong global prior for surface geometry and encourage a more coherent arrangement of Gaussian primitives, we jointly optimize a neural network representing an implicit SDF of the scene, $\bm{f_{neural}(x)}$, alongside the Gaussians. The SDF is represented by a Multi-Layer Perceptron (MLP), termed Implicit SDFNet in our model. As shown in \cref{fig:threecol}, this network takes 3D point coordinates as input and outputs a scalar SDF value. However, standard Multi-Layer Perceptrons tend to learn low-frequency variations in data much more easily than high-frequency variations. This is known as "spectral bias." To capture high-frequency details, input coordinates are first transformed using positional encoding (e.g., 6 frequency bands using sinusoidal functions).
$\sin(2^i \cdot x)$ and $\cos(2^i \cdot x)$ for different $i$.

The MLP consists of 8 hidden layers with 256 neurons each using a Softplus activation function. Skip connections are incorporated between the original encoded input to an intermediate layer's input to improve gradient flow and detail representation. The network weights are initialized as 0 with small negative value on output layer bias, similar to practices in \cite{author5} and ~\cite{author9}, to promote stable SDF learning.\\

\noindent
\textbf{Eikonal Loss:} To ensure that $\bm{f_{neural}(x)}$ behaves as a valid distance field, an Eikonal loss is applied. This loss penalizes deviations of the SDF's gradient norm from being 1: 
\begin{equation}
\mathcal{L}_{Eikonal} =\mathbb{E}_{x} \left( \left\| \bm{\nabla f_{neural}(x)} \right\|_2 - 1 \right)^2
\label{eq:five}
\end{equation}
\noindent
Points $x$ for this loss are sampled around Gaussians in the scene, and gradients are computed via automatic differentiation. This loss is weighted by $\lambda_{eikonal}$. \\

\noindent
\textbf{SDF-Gaussian Consistency Loss:}
We introduce To ensure a consistency loss so that the explicit Gaussians can align to the learned implicit surface. This loss encourages the regions represented by the Gaussian primitives to lie on or near the zero-level set of $\bm{f_{neural}(x)}$. This is achieved by sampling points $x_i$ based on the distribution of the current Gaussian primitives and penalizing the squared SDF values at these locations: 
\begin{equation}
\mathcal{L}_{\text{SDF-Gauss}} = \mathbb{E}_{x_i \sim \text{Gaussians}} \left( \bm{f_{neural}(x_i)} \right)^2
\label{eq:five}
\end{equation}

\noindent
Since the distance between sampled point and SDF surface gets penalized, the Gaussians and SDF will align as we iterate the process. This loss is weighted by $\lambda_{SDF-Gauss}$.

By integrating these regularization techniques, our methodology aims to produce Gaussians that are individually well-shaped and collectively define a more accurate and globally consistent surface. At the same time, learned SDF converges to object's true shape, making the Gaussians align more closely with the surface and thereby improving the rendering quality of the subsequently converted mesh.

\section{Experiment}
We used the Mip-NeRF 360 dataset for our experiments. First, we ran 7,000 iterations of vanilla Gaussian Splatting to allow the Gaussians to form a coarse shape. Then, for 8,000 iterations, we performed the Gaussian alignment process using our defined loss functions. The experiments were conducted on an RTX 4090 GPU, and mesh conversion took approximately 1.5 to 2 hours.

\subsection{Result}
In \cref{tab:tabone}, we show quantitative result comparing basic SuGaR model and our newly proposed model. The result show that our method shows better rendering quality in many different metrics.

\begin{table*}
\centering
\begin{center}
\begin{tabular} {ccccccccccc}
\hline\hline
& Method &\multicolumn{3}{c}{Stump}&&\multicolumn{3}{c}{Bonsai}&\\
\cline{3-5} \cline{7-9}
& & SSIM$\uparrow$ & PSNR$\uparrow$ & LPIPS$\downarrow$ & & SSIM$\uparrow$ & PSNR$\uparrow$ & LPIPS$\downarrow$ & \\
\hline
& 2DGS & 0.7885 & 26.5749 & 0.1948 && 0.9508 & 31.8650 & \textbf{0.0730} &\\
\hline
& SuGaR & 0.7657 & 26.8959 & 0.2056 && 0.9477 & 31.9473 & 0.0955 &\\
\hline
& Ours & \textbf{0.7927} & \textbf{27.2257} & \textbf{0.1922} && \textbf{0.9509} & \textbf{32.3401} & 0.0823 & \\
\hline\hline
& Method &\multicolumn{3}{c}{Counter}&&\multicolumn{3}{c}{Garden}&\\
\cline{3-5} \cline{7-9}
& & SSIM$\uparrow$ & PSNR$\uparrow$ & LPIPS$\downarrow$ & & SSIM$\uparrow$ & PSNR$\uparrow$ & LPIPS$\downarrow$ & \\
\hline
& 2DGS & 0.9102 & 28.9066& \textbf{0.0967} && 0.8906 & 28.0669 & 0.0855 &\\
\hline
& SuGaR & 0.9059 & 28.4876 & 0.1175 && 0.9124 & 29.5415 & 0.0823 &\\
\hline
& Ours & \textbf{0.9120} & \textbf{29.0136} & 0.1089 && \textbf{0.9252} & \textbf{30.3401} & \textbf{0.0783} & \\
\hline\hline
& Method &\multicolumn{3}{c}{Kitchen}&&\multicolumn{3}{c}{Room}&\\
\cline{3-5} \cline{7-9}
& & SSIM$\uparrow$ & PSNR$\uparrow$ & LPIPS$\downarrow$ & & SSIM$\uparrow$ & PSNR$\uparrow$ & LPIPS$\downarrow$ & \\
\hline
& 2DGS & 0.9468 & 31.1812 & \textbf{0.0477} && 0.9404 & 30.3269 & \textbf{0.1004} &\\
\hline
& SuGaR & 0.9462 & 31.5774 & 0.0587 && 0.9078 & 29.7830 & 0.1294 &\\
\hline
& Ours & \textbf{0.9524} & \textbf{31.9214} & 0.0496 && \textbf{0.9812} & \textbf{30.5670} & 0.1121 & \\
\hline\hline
\end{tabular}
\end{center}
\caption[Quantitative result]{Our method with advanced regularization show better performance than original SuGaR model.}
\label{tab:tabone}
\end{table*}

\begin{figure*}
  \centering
  \includegraphics[width=0.8\linewidth]{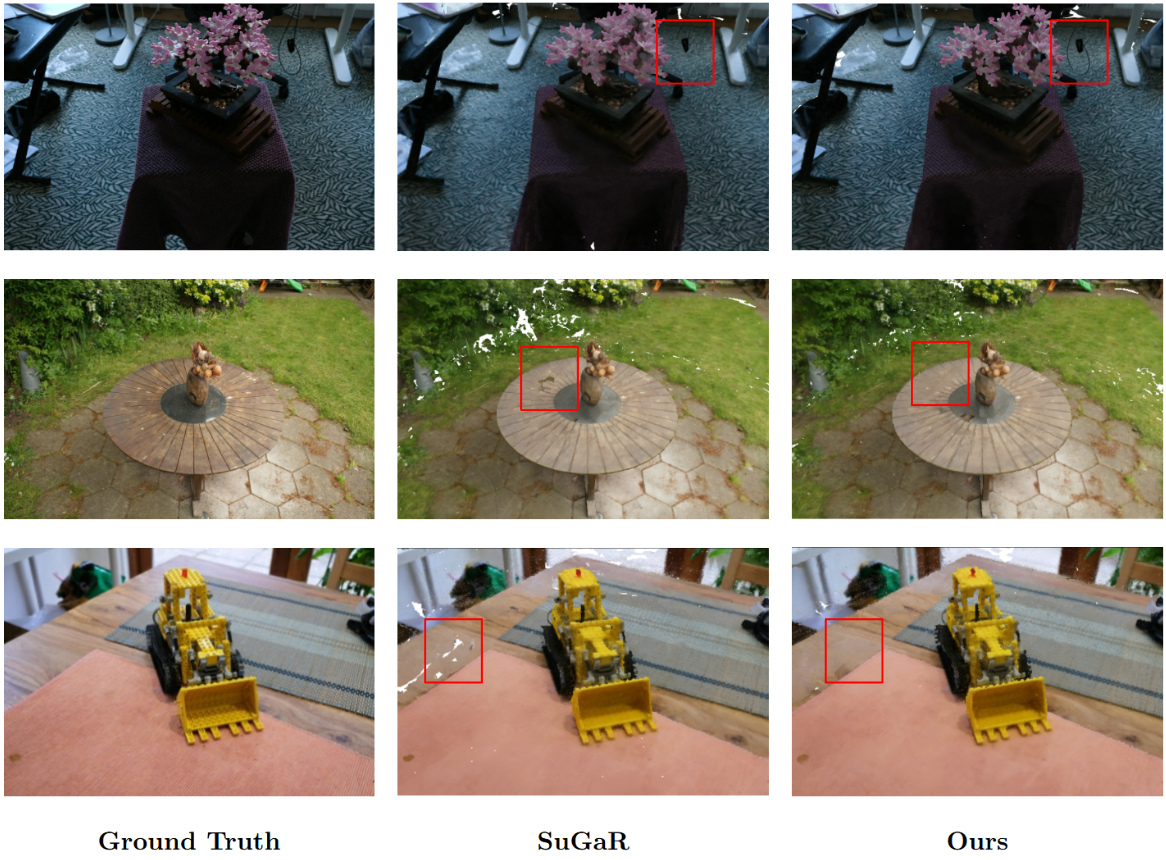}

   \caption{Qualitative mesh render comparison between SuGaR and our model.}
   \label{fig:fivecol}
\end{figure*}

We also show qualitative result with \cref{fig:fivecol} of actual mesh renderings on Mip-NeRF 360 dataset. The rendered images show that extracted mesh with our method covers more area(less white areas) and show better rendering. Also note that original SuGaR model cannot properly render the electric wire at upper right side of the image whereas our model does it better. Also our model doesn't have artifacts that original SuGaR model had as shown in the middle and bottom row. 

\begin{table*}
\centering
\begin{center}
\begin{tabular} {ccccccccccc}
\hline\hline
& Method &\multicolumn{3}{c}{Stump}&&\multicolumn{3}{c}{Bonsai}&\\
\cline{3-5} \cline{7-9}
& & SSIM$\uparrow$ & PSNR$\uparrow$ & LPIPS$\downarrow$ & & SSIM$\uparrow$ & PSNR$\uparrow$ & LPIPS$\downarrow$ & \\
\hline
& w/o Any regularization & 0.7657 & 26.8959 & 0.2056 && 0.9477 & 31.9473 & 0.0955 &\\
\hline
& w/o Effective Rank & 0.7792 & 27.0218 & 0.2001 && 0.9501 & 32.1174 & 0.0933 &\\
\hline
& w/o Implicit Net & 0.7649 & 26.9913 & 0.1948 && 0.9496 & 31.9870 & 0.0938 &\\
\hline
& Ours & \textbf{0.7927} & \textbf{27.2257} & \textbf{0.1922} && \textbf{0.9509} & \textbf{32.3401} & \textbf{0.0823} & \\
\hline\hline
& Method &\multicolumn{3}{c}{Counter}&&\multicolumn{3}{c}{Garden}&\\
\cline{3-5} \cline{7-9}
& & SSIM$\uparrow$ & PSNR$\uparrow$ & LPIPS$\downarrow$ & & SSIM$\uparrow$ & PSNR$\uparrow$ & LPIPS$\downarrow$ & \\
\hline
& w/o Any regularization & 0.9059 & 28.4876 & 0.1175 && 0.9124 & 29.5415 & 0.0823 &\\
\hline
& w/o Effective Rank & 0.9089 & 29.0072 & 0.1116 && 0.9188 & 29.8991 & 0.0801 &\\
\hline
& w/o Implicit Net & 0.9068 & 28.8721 & 0.1136 && 0.9209 & 30.0290 & 0.0798 &\\
\hline
& Ours & \textbf{0.9120} & \textbf{29.0136} & \textbf{0.1089} && \textbf{0.9252} & \textbf{30.3401} & \textbf{0.0783} & \\
\hline\hline
\end{tabular}
\end{center}
\caption[Ablation Study]{We get higher rendering quality by incorporating our regularization terms.}
\label{tab:tabtwo}
\end{table*}

\newpage

\subsection{Ablation Study}
In \cref{tab:tabtwo}, we demonstrate the effectiveness of our method. We have conducted an ablation study without using Effective Rank regularization and Implicit SDF regularization. We observed that including either effective rank or implicit net regularization slightly improves the overall performance. Including both regularizations further improves the rendering quality. The regularization steps add up to 20 minutes to the process, but we prioritized better rendering quality over the increased time.
\section{Future Work}
\noindent
While our proposed enhancements demonstrate promising mesh rendering quality from 3D Gaussian Splatting via the SuGaR framework, several limitations and avenues for future research exist: \\

\noindent
\textbf{Limitations from vanilla GS:} Since our approach heavily relies on the quality of the Gaussians fed into SuGaR’s mesh extraction pipeline, the limitations of vanilla Gaussian Splatting persist in our model. For instance, non-Lambertian surfaces like glossy or reflective materials may still show significant challenges. The effectiveness of the current regularization under such extreme conditions requires further evaluation. Recent works have explored combining Gaussian Splatting with ray tracing to address these issues~\cite{author18, author19}. Additionally, works like EnvGS~\cite{author20}focuses on learning spherical harmonics from multiple viewpoints. \\

\noindent
\textbf{Computational Overhead:} The introduction of a neural SDF network and the computation of additional loss terms inevitably add to the training time and computational cost compared to the original SuGaR pipeline or vanilla 3DGS. Optimizing the efficiency of these new components, perhaps by using more compact SDF network architectures or adaptive application of the losses, could be explored.\\

\noindent
\textbf{Mesh Extraction Method:} Our approach enhances the alignment of the Gaussians fed into SuGaR's mesh extraction pipeline, which relies on point sampling from a density level set and Poisson reconstruction. If the co-optimized neural SDF achieves very high fidelity, directly extracting the mesh from this SDF could be an alternative. This would represent a more significant departure from the SuGaR pipeline but might offer different trade-offs in terms of detail and smoothness.\\

\noindent
\textbf{Interaction with Refinement Stage:} Our current enhancements primarily focus on the alignment stage of the SuGaR pipeline. A more detailed investigation into how an improved coarse mesh impacts the subsequent optional refinement stage (where new Gaussians are bound to the mesh) could yield further insights and potential optimizations. \\

\noindent
\textbf{Gaussian Primitives:} We addressed the anisotropy caused by "needle-like" Gaussians through effective rank regularization, but a low-pass filtering tendency still persists. This can cause fine details to be blurry unless an extremely large number of Gaussians are used. DyGASR~\cite{author21} attempts to mitigate this issue by replacing Gaussians with a generalized exponential function for splatting. By using a higher exponent, the representation can capture sharper transitions, reducing the number of primitives needed to preserve crisp details. \\

In the future, these regularization techniques could be applied to other types of explicit 3D representations where maintaining temporal coherence of both primitive shapes and global surfaces is critical.
In this work, we introduced improvements to the SuGaR framework to enable more reliable 3D mesh rendering from 3DGS representations. While 3DGS excels at real-time rendering, turning its Gaussian primitives into clean, accurate surface meshes remains a difficult task. To address this, we incorporated two complementary regularization strategies during SuGaR’s initial Gaussian-aligning phase.

The first is an effective rank regularization, designed to improve the shape quality of individual Gaussians. It discourages thin, elongated "needle-like" primitives and instead promotes more compact, disk-shaped ones—better suited to capturing local surface geometry. The second is a joint optimization approach that incorporates a neural SDF. This SDF is trained alongside the Gaussians, constrained by an Eikonal loss and a consistency loss that links the SDF to the Gaussians. As a result, it acts as a smooth global prior, helping to guide the Gaussians toward a more coherent and unified surface.

Together, these two forms of regularization—one targeting local shape and the other encouraging global alignment—lead to Gaussians that are not only individually well-formed but also collectively produce a more accurate and consistent representation of the scene. This significantly improves the quality of the rendering, making it more visually coherent. These changes make downstream applications like editing, animation, and physical simulation much easier. Our goal is to bridge the gap between the fast rendering advantages of 3DGS and the practical demands of high-quality geometric assets in graphics workflows.
{
    \small
    \bibliographystyle{ieeenat_fullname}
    \bibliography{main}
}

\end{document}